\documentclass[sigconf,natbib]{acmart}

\renewcommand\footnotetextcopyrightpermission[1]{}
\usepackage{algorithm}
\usepackage{algorithmic}
\usepackage{multirow}
\usepackage{amsmath}
\DeclareMathOperator*{\argmin}{arg\,min}

\AtBeginDocument{%
  \providecommand\BibTeX{{%
    \normalfont B\kern-0.5em{\scshape i\kern-0.25em b}\kern-0.8em\TeX}}}

\setcopyright{none}

\acmDOI{XXXXXXX.XXXXXXX}
\acmConference[]{Workshop}{ECI}{@CIKM2023}
\acmBooktitle{}
\acmPrice{}
\acmISBN{}
\acmSubmissionID{}

\begin{document}

\title{Search Optimization with Query Likelihood Boosting and Two-Level Approximate Search for Edge Devices}

\author{Jianwei Zhang}
\email{jianwei.zhang@asu.edu}
\orcid{0000-0001-6419-2038}
\affiliation{%
  \institution{Arizona State University}
  \city{Phoenix}
  \state{Arizona}
  \country{USA}
  \postcode{85281}
}

\author{Helian Feng}
\email{hlfeng@amazon.com}
\affiliation{%
	\institution{Amazon Alexa AI}
	\city{Boston}
	\state{Massachusetts}
	\country{USA}
	\postcode{02142}}

\author{Xin He}
\email{xih@amazon.com}
\affiliation{%
	\institution{Amazon Alexa AI}
	\city{Boston}
	\state{Massachusetts}
	\country{USA}
	\postcode{02142}}

\author{Grant P. Strimel}
\email{gsstrime@amazon.com}
\affiliation{%
	\institution{Amazon Alexa Speech}
	\city{Pittsburgh}
	\state{Pennsylvania}
	\country{USA}
	\postcode{15203}}
	
\author{Farhad Ghassemi}
\email{gfarhad@amazon.com}
\affiliation{%
	\institution{Amazon Alexa AI}
	\city{Pittsburgh}
	\state{Pennsylvania}
	\country{USA}
	\postcode{15203}}

\author{Ali Kebarighotbi}
\email{alikeba@amazon.com}
\affiliation{%
  \institution{Amazon Alexa AI}
  \city{Boston}
  \state{Massachusetts}
  \country{USA}
  \postcode{02142}}

\renewcommand{\shortauthors}{J. Zhang et al.}

\begin{abstract}
We present a novel search optimization solution for approximate nearest neighbor (ANN) search on resource-constrained edge devices. Traditional ANN approaches fall short in meeting the specific demands of real-world scenarios, e.g., skewed query likelihood distribution and search on large-scale indices with a low latency and small footprint. To address these limitations, we introduce two key components: a Query Likelihood Boosted Tree (QLBT) to optimize average search latency for frequently used small datasets, and a two-level approximate search algorithm to enable efficient retrieval with large datasets on edge devices. We perform thorough evaluation on simulated and real data and demonstrate QLBT can significantly reduce latency by 15\% on real data and our two-level search algorithm successfully achieve deployable accuracy and latency on a 10 million dataset for edge devices. In addition, we provide a comprehensive protocol for configuring and optimizing on-device search algorithm through extensive empirical studies.
\end{abstract}



\maketitle

\section{Introduction}
As low-power edge devices like speakers and watches become increasingly integral to daily life, the need for locally executing traditional server-side tasks rises. Entity Resolution (ER), a key component for digital voice assistants (VAs) like Amazon Alexa and Apple Siri, is one of such tasks. ER maps entity mentions (i.e., query) to IDs/titles presented in a single or multiple datasets~\cite{getoor2012entity}. Approximate nearest neighbor (ANN) search algorithms are used for ER, which can identify and return the datapoints within a dataset that are the nearest to a query data point. Currently, most production ER applications rely on cloud resources with much larger memories and powerful processors than local devices to support advanced ANN methods. The increasing prevalence of VAs calls for local execution to improve latency and robustness against network connectivity and better address daily personalized consumer needs and data privacy~\cite{PanasonicAlexa2018CarVoiceControl}. Therefore, resolving the local ER challenge, constrained by limited computational resources, becomes crucial for ensuring a swift and consistent customer experience.

McGowan et al. proposed SmallER~\cite{mcgowan2021smaller} to address several challenges of deploying ER on devices: limited CPU MIPS and run-time memory, capped on-device storage space for the index and search solutions, and strict query latency requirements. They utilize a compact encoded spatial partitioning projection tree (SPPT) for ANN search, which enables on-device ANN search for small to medium datasets. The SPPT is a balanced binary tree to provide equal latency (constant depth) for all queries. However, the query likelihood distribution of real-world traffic is usually highly-skewed, i.e., fathead vs. long tail~\cite{kritzinger2013search}, with a small number of frequently queried entities. Although SPPT is superior to other classic ANN methods for on-device search for small to medium datasets, the balanced structure leads to sub-optimal average latency. Furthermore, SPPT tends to have high latency on large datasets. Similarly, other classic ANN methods, such as locality sensitive hash and product quantization, cannot meet the accuracy or latency requirement either.

We present a novel solution for ANN search on resource-constrained edge devices. We addressed the limitations for traditional ANN algorithm in real world scenario with two key components: (1) a Query Likelihood Boosted Tree (QLBT) to optimize average search latency for frequently used small datasets; and (2) a two-level approximate search to enable efficient retrieval with large datasets on edge devices. We perform thorough evaluation on simulated and real data and demonstrate QLBT can significantly reduce latency by 15\% on real data and our two-level search successfully achieve deployable accuracy and latency on a 10M dataset for edge devices. In addition, we provide a comprehensive protocol for configuring on-device ANN search through extensive empirical studies.

\section{Related work}

We briefly review classic ANN methods commonly used for ER. To improve search speed and reduce memory usage, these algorithms mildly relax accuracy constraints~\cite{ebraheem2018distributed, gillick2019learning}. Some graphic-methods, like hierarchical navigable small world~\cite{malkov2018efficient} and others~\cite{zhao2021approximate, peng2022lan}, are excluded from our scope due to the on-device storage limitation.

\textbf{Spatial Partitioning Tree} Tree-based search structures for ANN have many variations including kd-tree~\cite{bentley1975multidimensional}, R-tree~\cite{guttman1984r}, random projection tree~\cite{dasgupta2008random}, etc. The core idea of these implementations is to partition the data into recursively divided regions represented in a tree structure. Similar neighbors are assumed to reside under a closer hierarchy of the resulting tree. While tree-based ANN performs well when the dataset is small, often when the dataset is large, extensive probing of many neighboring branches becomes unavoidable~\cite{wang2022graph}, leading to untenable search time.

\textbf{Locality Sensitive Hash (LSH)} LSH uses hash functions to allow similar keys or vectors stay in the same bucket with a higher likelihood.  LSH is attractive because of its simplicity and fast search speed and simple random projection hash functions~\cite{charikar2002similarity} is often effective. However, hyperparameters tuning for LSH is known to be notoriously difficult~\cite{cai2019revisit}. Thus, empirically tree-based methods have achieved generally better recall than LHS for ANN tasks on standardized datasets~\cite{sinha2014lsh}.

\textbf{Product Quantization (PQ)} PQ effectively encodes high dimensional vectors by an optimized sub-vector space codebook for storage compression and guide fast ANN search~\cite{ge2013optimized}. The distance between two vectors can be approximated by the distance between their codewords, which is faster than Euclidean distance computing. IndexIVFPQ method in FAISS is a SOTA ANN algorithm which uses PQ for sub-datasets searching~\cite{jegou2011searching}. However, PQ can be computational intensive on large dataset~\cite{wang2022graph}.

These existing ANN methods all fall short to meet the specific demands of real-world Edge device scenarios, specifically, the skewed query likelihood distribution and search on large dataset under low latency and footprint. In Section~\ref{sec:method}, we design a ER system to bridge the aforementioned gaps with a QLBT, a flexible two-level search and a comprehensive on-device search optimization protocol.

\section{Methodology} \label{sec:method}

\subsection{Query likelihood boosted tree}

VA ER systems' traffic usually has a heavily skewed frequency distribution with frequent queries for a few entities and rare visits for the rest, forming a long tail. The traditional ANN implementations (excluding augmentations like caching mechanisms) are designed to have approximately the same latency for each entity without factoring in query likelihood. We exploit the skewed frequency distribution to our advantage, adjusting for rare entities, to enhance average latency and optimize the experience for head traffic.

Inspired by the source coding techniques~\cite{gray2011entropy}, we aim to minimize the anticipated depth of an entity in the projection tree to reduce the average latency during a query search:

\begin{equation*}
	\arg\min_{\theta} E[\text{Depth}(X;\theta)] = \arg\min_{\theta}\sum_{i}p(x_{i})\text{Depth}(x_{i};\theta),
\end{equation*}

\noindent where $x_{i}\in X$ is an entity, $\theta$ is the projection tree building rule, ${Depth}(x_{i};\theta)$ is the depth of $x_{i}$ in a projection tree built under rule $\theta$, and $p(x_{i})$ is the query likelihood for $x_i$.

We integrate the Shannon-Fano coding~\cite{jones2012information} with a randomized SPPT~\cite{mcgowan2021smaller} to develop the QLBT. Our primary innovation lies in our splitting criteria where the splitting hyperplane is determined based on an equal probability of visiting the left and right branches. When the number of entities in the left and right branches are unbalanced (entities number in one branch is much smaller than another branch), the entities with higher query likelihood will be positioned closer to the root node, i.e. shallower depth and consequently reduced search time. Beside the novel splitting criteria, we also factor in the variability of the data projected along the corresponding projection to determine the best projection hyperplane. 

To hedge the risk that unbalanced trees may have large latency on tail traffic (rare entities) and allow effective exhaustive searches on bottom leaves, we apply additional regulations on the QLBT building procedure for robustness: \textbf{(1) early stopping on query likelihood boosting} - once the partitioning tree building reach depth $\ell$ (we use $\ell=3$), stop balancing on query likelihood (i.e., roll back to the balanced tree); \textbf{(2) pre-grouping leaves} - we pre-group leaves during building, i.e., maintain multiple objects at the leaves instead of a single entity. In our implementation, 8 entities are grouped into one leaf for fast retrieval. Experiments show pre-grouping leaves reduce the latency without affecting recall~\cite{mcgowan2021smaller}. 

The recursive building procedure of QLBT is shown in Algorithm~\ref{qlb-tree-alg}. We denote the entity embeddings in dataset as $\mathbf{e_1, e_2 ...e_N}$ where each $\mathbf{e_i}$ is a $d$-dimension vector, and entity query likelihood in real traffic as $p_1,p_2,...p_N$. $\lambda$ is a hyper-parameter to control the trade-off between query likelihood unbalance level and data separation which we performed grid search for optimization.

\begin{algorithm}
	\caption{Query likelihood boosted tree building procedure}\label{qlb-tree-alg}
	\begin{algorithmic}[1]
	\REQUIRE Entity embedding $\mathbf{e_1...e_m}$ of current node, current node depth $D$, early-stop boosting level $\ell$, trade-off parameter $\lambda$
	\IF{$m \leq $ max leaf size}
		\STATE Save inference IDs of entities, return
	\ELSE{}
		\STATE Generate $K$ random projection $\mathbf{v_{1}, ..., v_{K}}$ on unit sphere
		\FOR {Each projection vector $\mathbf{v_i}$}
			\STATE Calculate projections: $\alpha_{i,j} = \mathbf{v_i\cdot e_j}$
			\STATE Find $\tau^* = \argmin_\tau |\sum_{\alpha_{i,j} \leq \tau} p(x_j) - \sum_{\alpha_{i,j} > \tau} p(x_j)|$
			\STATE $N_\text{left} = |\left\{ \alpha_{i,j} \leq \tau^* \right\}|, N_\text{right} = |\left\{ \alpha_{i,j} > \tau^* \right\}|$
			\STATE Unbalance score: $b_{i} = \max(N_\text{left} / N_\text{right}, N_\text{right} / N_\text{left})$
			\STATE Projection variations: $\sigma^{2}_{i} = Var(\alpha_{i,1}, \dots , \alpha_{i,m})$
			\IF {$D \leq \ell$}
				\STATE Projection score of $\mathbf{v_i}$: $score_{i} = \lambda\cdot\sigma^{2}_{i} + (1-\lambda)\cdot b_{i}$
			\ELSE{}
				\STATE Projection score of $\mathbf{v_i}$: $score_{i} = \sigma^{2}_{i}$
			\ENDIF
		\ENDFOR
		\STATE Best projection vector $\mathbf{v_{best}} \gets \max_\mathbf{v_i}score_{i}$
		\STATE Split the entities of current node into left and right child-nodes with $\mathbf{v_{best}}$ and corresponding $\tau^*$
		\STATE Repeat algorithm on left and right child-nodes
	\ENDIF
	\end{algorithmic}
\end{algorithm}

Though a different construction, we use the same searching procedure described in~\cite{mcgowan2021smaller} for QLBT. We emphasize that the boosted search tree must be updated for changes in users' query likelihood distribution. If only this distribution changes, a new search tree can be easily built, keeping other configurations the same. This method simplifies updates and enables personalized search experiences.

\subsection{Two-level approximate search}

Traditional ANN approaches such as projection trees, LSH, and PQ often fail to search large-scale datasets on edge devices with required low latency and small footprint. One effective strategy to improve search efficiency is pre-partitioning the entire dataset into several smaller subsets. These rules can be derived from the entity embedding distribution or other natural heuristics such as metadata, geolocation. After partitioning, appropriate search algorithms can be applied to each subset individually. Pre-partitioning is also a common strategy employed to search large datasets in SOTA ANN algorithms, like IndexIVFPQ in FAISS~\cite{jegou2011searching}.

We propose a two-level approximate search for edge ER, which is shown in Figure~\ref{fig:grid}(a): (1) derive dataset pre-partitioning rules from features, e.g., vectorized entity embeddings, geolocation; (2) run K-means clustering on the partitioning features to create sub-datasets with corresponding centroids; (3) index the top-level on these centroids and complete ANN search based on entitiy embeddings over entities on the bottom-level. Our proposed method stems from the integration of top and bottom-level search mechanisms, enhancing efficiency over traditional methods. Utilizing feature-driven partitioning and K-means clustering, it can manage larger indices with lower latency, making it ideal for edge devices.

We offer a collection of choices for top- and bottom-level algorithms. We consider three algorithms for the top-level: (1) brute search, used when sub-datasets number is small; (2) kd-tree~\cite{bentley1975multidimensional}, used when the pre-partitioning features have low dimensions (e.g., latitude and longitude geolocation); (3) PQ~\cite{jegou2010product}, used when the pre-partitioning features have large dimensions (e.g. embedding). For the bottom-level we consider: (1) brute search, (2) QLBT, and (3) footprint-reduced LSH via a fixed set of random projections.

\section{Experiment Setup}

To fairly compare different ANN search methods' performance on private and public datasets, we implement an ANN benchmark on an AWS t3.xlarge instance (3.1 GHz Intel Xeon processor, 16GB memory) based on an open-source ANN benchmark~\cite{aumuller2020ann}. It provides a standardized way to evaluate different ANN algorithms. By setting a common ground for comparison, it allows us to understand how different algorithms perform under various scenarios, datasets, and parameters. All results are generated by this benchmark. Two searching limits are set to reflect the customer requirement for Edge ER: (1) P90 time less than 80ms with Python implementation; (2) recall@10 more than 80\%, where recall@$k$ is the percent of test dataset with the ground truth entity among top $k$ returned entities.

\subsection{Datasets}

To comprehensively evaluate our on-device ANN search framework, we conduct experiments on a series of private and public ANN datasets with dataset sizes ranging from 10K to 10M. We use:

\textbf{Radio Station}: private, test sets sampled from real traffic, contains around 10K entities of 256d vectors.

\textbf{SIFT}: public, contains 1M entities of 128d vectors~\cite{jegou2010product}.

\textbf{DEEP1B}: public, contains 10M entities of 96d vectors~\cite{babenko2016efficient}.

\subsection{Query likelihood simulation}\label{sec:qlb-sim}

To broadly assess the latency reduction with the proposed QLBT on various query likelihood distribution, we simulate different query likelihood distribution for 256 entities from Radio Station dataset via a Beta distribution. Each simulated dataset contains 10K queries sampled from Radio Station dataset. We use a query likelihood unbalancing score (based on information entropy~\cite{shannon2001mathematical}) to describe the likelihood unbalancing level of these entities: $1 - \frac{-\sum_i^N p(x_i) \text{log}_2 p(x_i)}{ \text{log}_2 N}$, where the $N$ is the number of entities. High unbalance score denotes high skew level of the query likelihood distribution and a uniform distribution has a zero unbalance score. Our real world radio station query traffic has an unbalance score of $0.23$.

\section{Results}

\subsection{Query likelihood boosted tree search}

\begin{figure}[ht]
	\centering
	\includegraphics[width=0.85\linewidth]{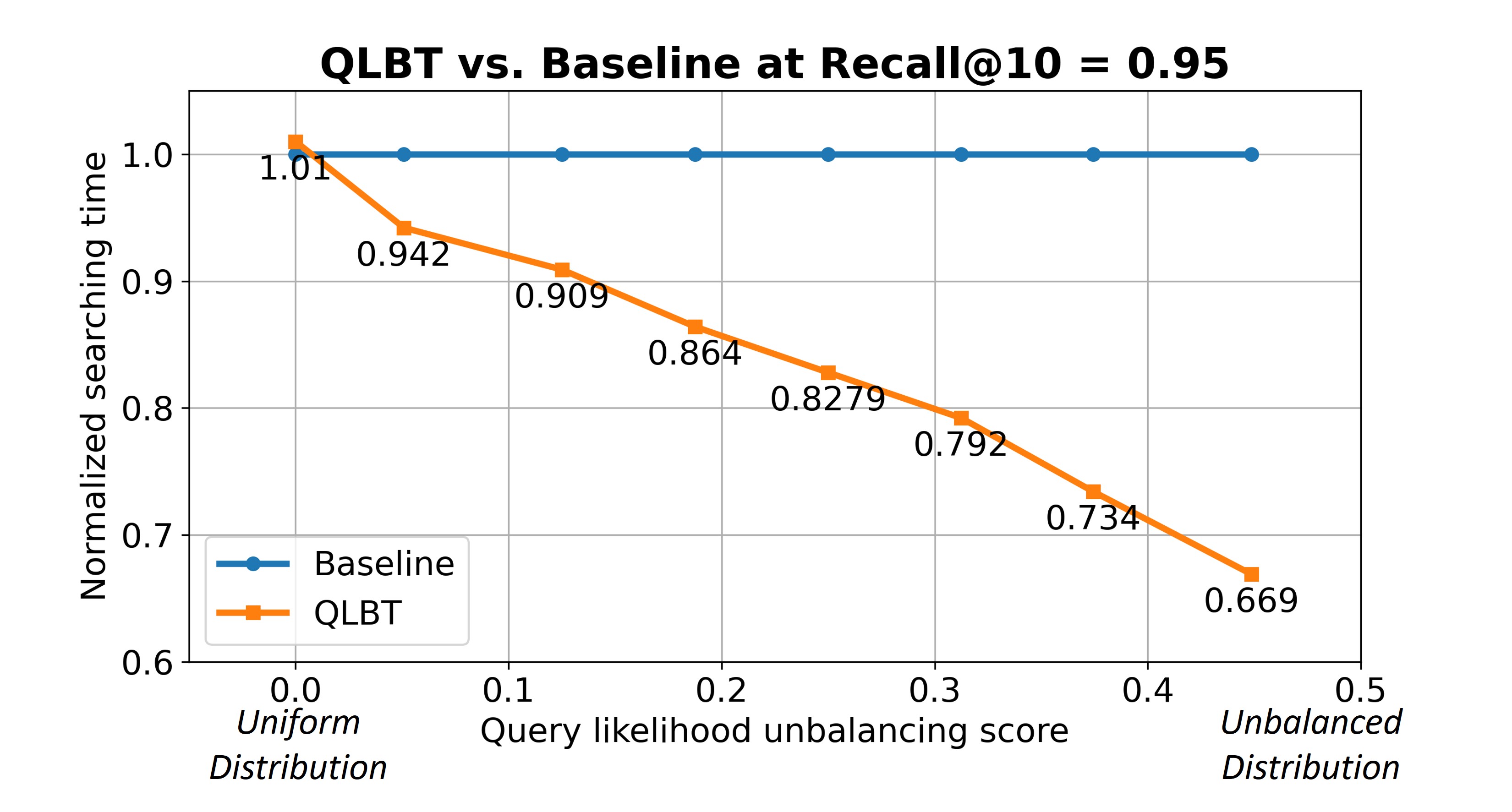}
	\caption{P90 time (recall@10 = 0.95) changes vs. query likelihood unbalancing score on QLBT and baseline tree.}
	\label{fig:simu}
\end{figure}

Using the simulation dataset described above, Figure~\ref{fig:simu} plots the relative search latency gain of QLBT as the unbalance score increases. We measure the P90 latency to achieve $0.95$ recall@10 and benchmark against $0.95$ recall@10 P90 latency of naive projection tree as baseline. Recall@$k$ is defined as the portion of test dataset which has their ground truths among the top $k$ returned entities. The results demonstrate that latency gain from QLBT increases with the unbalance level of the candidate query likelihood distribution. 

We further validate our finding on real-world Radio Station traffic data which has an unbalancing score of 0.23. Applying QLBT, we achieve $16\%$ latency reduction on P90 search time compared to naive projection tree at 0.95 Recall\@10 and $16\%$ reduction on the average search time. This result aligns well with the results on simulated data (Figure~\ref{fig:simu}), where an approximate $15\%$ latency reduction by QLBT are observed on data of 0.23 unbalance score. 

\begin{figure*}[!htb]
   \begin{minipage}{0.65\textwidth}
     \centering
     \includegraphics[width=0.85\linewidth]{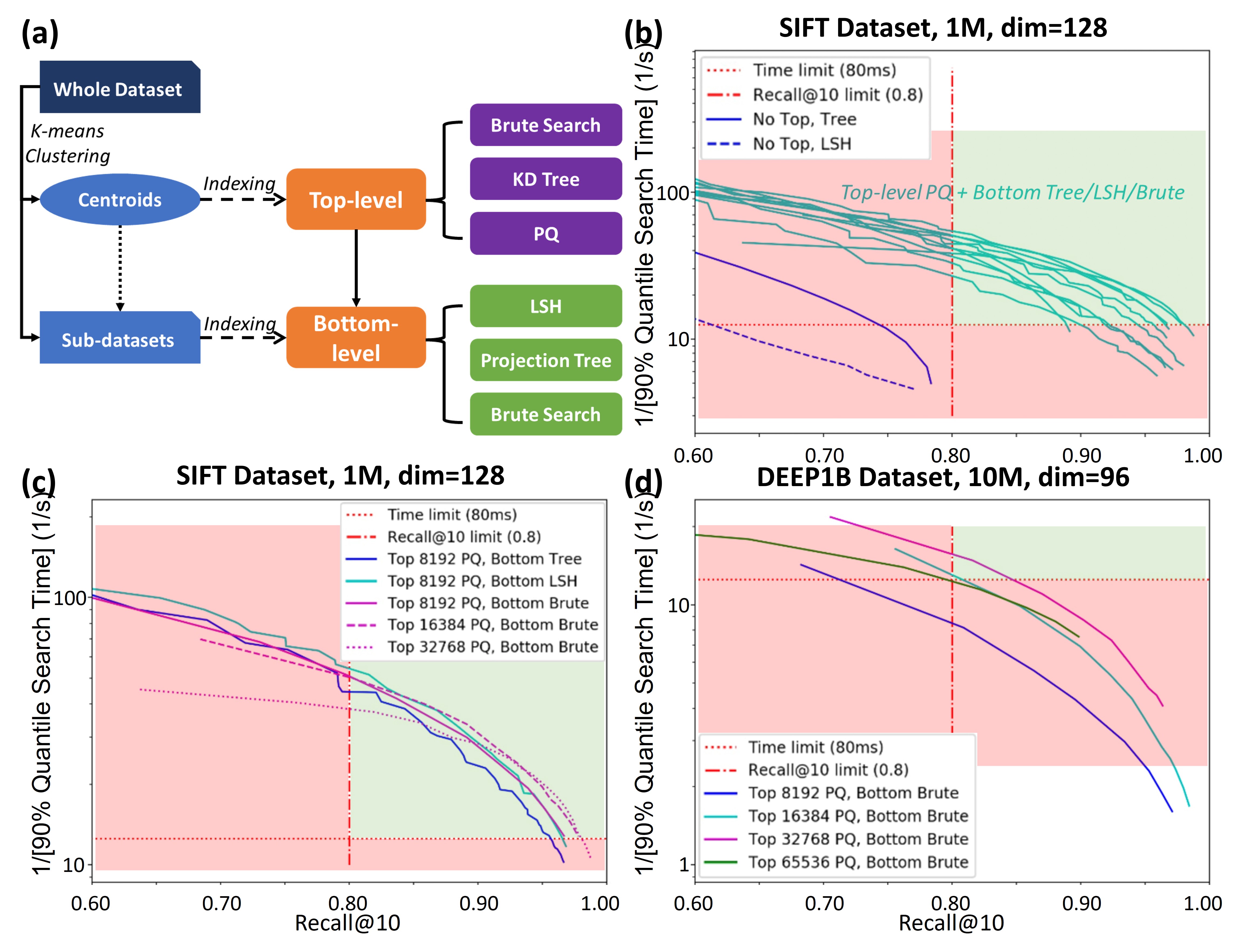}
     \caption{(a) Diagram for two-level search. (b) Pure tree and LSH vs. top PQ with $2^{10}$ to $2^{13}$ sub-datasets, bottom Tree/LSH/Brute. (c) Top PQ with $2^{13}$ and $2^{14}$ sub-datasets + bottom Tree/LSH/Brute on SIFT. (d) Top PQ + bottom Brute on DEEP1B.}
     \label{fig:grid}
   \end{minipage}\hfill
   \begin{minipage}{0.33\textwidth}
     \centering
     \includegraphics[width=1.1\linewidth]{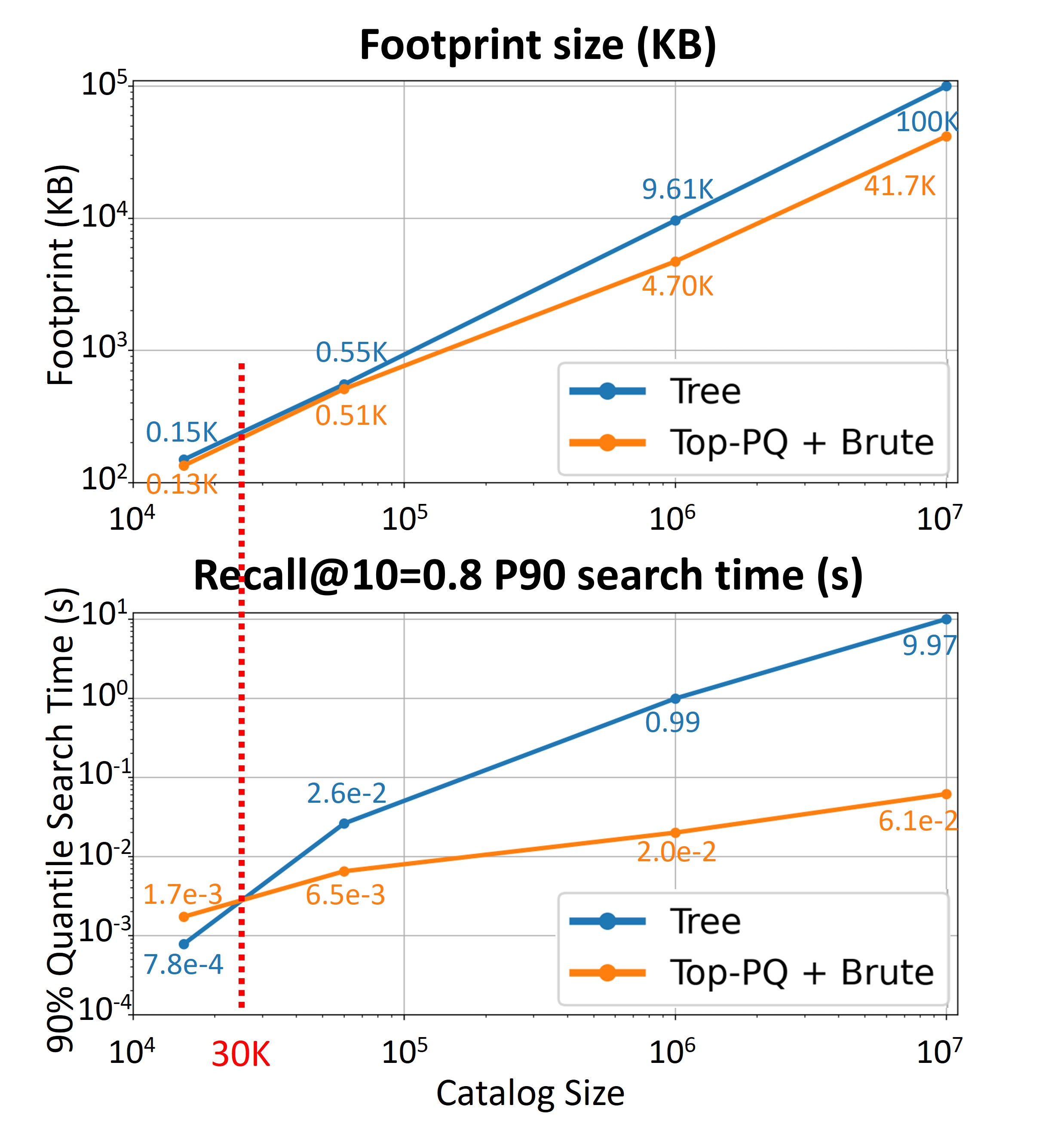}
     \caption{Footprint and P90 time comparison between one-level and two-level structure for catalogs with different sizes.}\label{fig:line}
   \end{minipage}
\end{figure*}


\begin{table*}[ht]
	\centering
	\caption{Recall@10 when P90 time = 80ms for experiments results on SIFT.}
	\label{tab:sift-res}
	\begin{tabular}{c|cc|ccccccccccc}
		\hline
		Config & \begin{tabular}[c]{@{}c@{}}No Top\\ /Tree\end{tabular} & \begin{tabular}[c]{@{}c@{}}No Top\\ /LSH\end{tabular} & \begin{tabular}[c]{@{}c@{}}PQ-$2^{10}$\\ /Tree\end{tabular} & \begin{tabular}[c]{@{}c@{}}PQ-$2^{10}$\\ /LSH\end{tabular} & \begin{tabular}[c]{@{}c@{}}PQ-$2^{11}$\\ /Tree\end{tabular} & \begin{tabular}[c]{@{}c@{}}PQ-$2^{11}$\\ /LSH\end{tabular} & \begin{tabular}[c]{@{}c@{}}PQ-$2^{12}$\\ /Tree\end{tabular} & \begin{tabular}[c]{@{}c@{}}PQ-$2^{12}$\\ /LSH\end{tabular} & \begin{tabular}[c]{@{}c@{}}PQ-$2^{13}$\\ /Tree\end{tabular} & \begin{tabular}[c]{@{}c@{}}PQ-$2^{13}$\\ /LSH\end{tabular} & \begin{tabular}[c]{@{}c@{}}PQ-$2^{13}$\\ /Brute\end{tabular} & \begin{tabular}[c]{@{}c@{}}PQ-$2^{14}$\\ /Brute\end{tabular} & \begin{tabular}[c]{@{}c@{}}PQ-$2^{15}$\\ /Brute\end{tabular} \\ \hline
		Recall & \color{red}{0.743~$\times$}                                                  & \color{red}{0.613~$\times$}                                                 & 0.899                                                & 0.922                                               & 0.919                                                & 0.905                                               & 0.923                                                & 0.944                                               & 0.957                                                & 0.966                                               & 0.968                                                 & 0.979                                                  & \color{teal}{\underline{0.980}\checkmark}                                                  \\ \hline
	\end{tabular}
\end{table*}


\subsection{Two-level approximate search}\label{sec:two-level}

We evaluate two-level search on two large scale retrieval datasets with more than 1M entities (SWIFT and DEEP1B) and produce the performance curve of recall\@10 vs 1/P90 latency of various search configurations. We record the best recall achieved on the given P90 constraint. Two searching limits are set as described in Section 4. In the performance curve plots (Figure~\ref{fig:grid} (b), (c), (d)), the algorithm appear on the most top-right has the best performance and only the curves fall in the green area meet the two searching limits.

Figure~\ref{fig:grid}(b) shows the SIFT data experimental results. The one-level methods (tree and LSH) performance are shown in blue lines and two-level search algorithm with various configurations are in green lines. We also provide the recall@10 at 80ms P90 search time in Table~\ref{tab:sift-res}). Neither of the one-level methods are able to achieve satisfactory recall at required latency. Whereas various two-level search configuration reach the acceptable performance range, with the best recall@10 at 0.980 from the top PQ + bottom Brute search configuration with $2^{15}$ sub-dataset split. The experiment result  prove the necessity of a two-level search algorithm for large dataset.

From Figure~\ref{fig:grid}(b) and Table~\ref{tab:sift-res}, we observe that the performance increases with the number of sub-datasets and brute force as the best performing bottom method. We increase the sub-datasets number to $2^{13}$ and $2^{14}$ and the result is shown in Figure~\ref{fig:grid}(c). When the the average entities number within each subset is around 100, the two-level search achieves the optimal, and we notice again that brute search on the bottom-level outperforms ANN search methods (e.g., tree and LSH). In conclusion, two-level ANN search achieves optimal performance with PQ as the top-level search algorithm and brute search as the bottom-level and the average entities number in subset is around 100. The conclusion is also validated on the bigger 10M DEEP1B datased shown in Figure~\ref{fig:grid}(d).

\subsection{Edge ER Configuration Optimization}

To provide guidance on optimal algorithm to generalize in more scenarios, we run experiments with different dataset sizes. The results, shown in Figure~\ref{fig:line}, reveal that footprint for one-level tree search and two-level search are similar for dataset size below 100K and two-level search has supreme P90 compared to one-level tree search when dataset size beyond 30K. Thus, we provide the guideline for on-device ANN search algorithm selection as below:

\noindent \textbf{Dataset size is smaller than 30K:}

\noindent (1) Traffic distribution available $\Rightarrow$ likelihood boosted tree

\noindent (2) Traffic distribution not available $\Rightarrow$ standard projection tree

\noindent \textbf{Dataset size is larger than 30K:}

\noindent (1) Top-level partitioning feature is high dimension (e.g., embedding) $\Rightarrow$ two-level approximate search with PQ + brute search and average subsets entities number is around 100

\noindent (2) Top level partitioning feature is low dimension (e.g., geo-location) $\Rightarrow$ two-level approximate search with top kd-tree and decide the bottom algorithm:

(a) If average subsets entities number $\leq$ than 100 $\Rightarrow$ brute

(b) If average subsets entities number $>$ 100 $\Rightarrow$ tree

\section{Conclusion}

We propose two optimized ANN search algorithms for Edge ER, the query likelihood boosted tree and two-level approximate search. The QLBT reduces the average real world query latency by 16\%. The two-level approximate search enables on-device ANN search with a 10M dataset, which is unachievable by previous methods. We also provide protocols for the best configuration of on-device search algorithm. For future work, we will implement the proposed algorithms on real edge devices and intend to include personalization into the on-device search.

\begin{acks}
We thank Zhimin Peng for providing Local Search data and Zimeng (Chris) Qiu for providing Audible data and Jiyang Wang, Nicholas Dronen for helpful discussions on the topics covered.
\end{acks}

\bibliographystyle{ACM-Reference-Format}
\bibliography{ref}



\end{document}